\documentclass[11pt]{article}
\usepackage{type1cm}
\usepackage[dvipdfmx]{graphicx}
\usepackage{cite}

\usepackage{amsmath,amssymb,bm,amsthm,cases}

\def\beq{\begin{equation}}
\def\eeq{\end{equation}}
\def\nbeq{\begin{equation*}}
\def\neeq{\end{equation*}}

\def\<{\langle}
\def\>{\rangle}

\renewcommand{\d}{\partial}

\newtheorem{theorem}{Theorem}
\newtheorem{lemma}{Lemma}

\begin{document}
\title{Existence of shape-dependent thermodynamic limit in spin systems with short- and long-range interactions}
\author{Takashi Mori \\
{\it
Department of Physics, Graduate School of Science,} \\
{\it The University of Tokyo, Bunkyo-ku, Tokyo 113-0033, Japan
}}
\maketitle

\begin{abstract}
The existence of the thermodynamic limit in spin systems with short- and long-range interactions is established.
We consider the infinite-volume limit with a fixed shape of the system.
The variational expressions of the entropy density and the free energy density are obtained, which explicitly depend on the shape of the system.
This shape dependence of thermodynamic functions implies the nonadditivity, which is one of the most important characteristics of long-range interacting systems.
\end{abstract}

\section{Introduction}
\label{sec:intro}

The aim of statistical mechanics is extracting thermodynamic properties from microscopic Hamiltonian.
Some thermodynamic properties and macroscopic phenomena can be well described by taking the thermodynamic limit~\cite{Ruelle_text,Griffiths_review1972}.
For example, thermodynamic quantities do not fluctuate in thermodynamics, which is exactly true only in the thermodynamic limit from the microscopic point of view.
A thermodynamic system sometimes exhibits a phase transition, which is well characterized as a mathematical singularity only in the thermodynamic limit.
Actually the system of interest is always finite, and thus the thermodynamic limit should be regarded as a theoretical idealization to extract thermodynamic properties from a given Hamiltonian.

From the statistical-mechanical point of view, it is a problem whether such a thermodynamic limit exists.
In short-range interacting systems, existence of the thermodynamic limit is well established, see~\cite{Ruelle_text,Griffiths_review1972}.
There, thermodynamic functions in the thermodynamic limit are shown to have appropriate convexity or concavity consistent with thermodynamics.
While, in long-range interacting systems, the existence of the thermodynamic limit has not been shown rigorously with sufficient generality.
Since many works reveal the peculiarities of long-range interacting systems~\cite{Campa_review2009,Les_Houches2009} such as the ensemble inequivalence and the negative specific heat, it is important to show the existence of the thermodynamic limit rigorously for general cases, e.g. the interaction potential is arbitrary under some natural conditions and the shape of the system is arbitrary.

In this paper, we shall establish the thermodynamic limit of classical spin systems with short- and long-range pair interactions satisfying some natural conditions specified later for arbitrary spacial dimension $d$ and arbitrary shape of the system specified by $\gamma$, see Sec.~\ref{sec:setup}.
We also obtain the variational expression of the entropy density in the thermodynamic limit, which explicitly shows that the entropy density depends on $\gamma$ even in the thermodynamic limit.
This dependence on the shape of the system implies the lack of additivity~\cite{Mori2014_quasi}, which is one of the most important characteristics of long-range interacting systems.

This paper is organized as follows.
In Sec.~\ref{sec:setup}, the setup and the notation are explained.
In Sec.~\ref{sec:theorem}, we mention the main result of this work, the existence of shape-dependent thermodynamic limit and the variational expression of the entropy density.
In Sec.~\ref{sec:proof} the proof is given.
In Sec.~\ref{sec:periodic}, we discuss the result of the derived variational form of the entropy density in the case of periodic boundary conditions.
In Sec.~\ref{sec:discussion}, we conclude this work and discuss a future prospect.

\section{Setup}
\label{sec:setup}

Let $\Gamma\subset\mathbb{R}^d$ be a bounded domain with volume $|\Gamma|$ on the $d$-dimensional space and $\hat{\Gamma}=\Gamma\cap\mathbb{Z}^d$ be the set of lattice points in $\Gamma$.
The number of elements of $\hat{\Gamma}$ is denoted by $N_{\Gamma}$.
We consider a classical spin system put on $\hat{\Gamma}$.
Each lattice point $\bm{r}\in\hat{\Gamma}$ has a spin variable $\sigma(\bm{r})$,
where $\sigma(\bm{r})$ may be a scalar or a vector.
The set of all the possible values of a spin variable is denoted by $\mathcal{S}$.
Here we assume that $\mathcal{S}$ is identical for all $\bm{r}\in\hat{\Gamma}$.
The set of $\sigma(\bm{r})$ for all $\bm{r}\in\hat{\Gamma}$ is denoted by $\bm{\sigma}_{\Gamma}\in\mathcal{S}^{N_{\Gamma}}$.

For simplicity, we consider the case in which $\sigma(\bm{r})$ is a scalar variable, $\mathcal{S}\subset\mathbb{R}$, in this paper, but the generalization to vector variables, $\mathcal{S}\subset\mathbb{R}^n$, where $n$ is the number of components of a spin variable, is straightforward.
Without loss of generality, we can assume $0\in\mathcal{S}$.
It is assumed that spin variables are bounded, $|\sigma(\bm{r})|\leq\sigma_{\rm max}$, where $\sigma_{\rm max}$ is independent of $\Gamma$.
Furthermore, we assume that the ``number of elements'' of $\mathcal{S}$ is finite,
$\sum_{\sigma\in\mathcal{S}}1=w<+\infty$.
For a continuous spin, $\sum_{\sigma\in\mathcal{S}}$ should be interpreted as $\int_{\mathcal{S}}\eta(\sigma)d\sigma$, where $\eta(\sigma)\geq 0$ is the weight of a state $\sigma$.

We consider the system described by the following Hamiltonian,
\beq
H_{\Gamma} = H_{\Gamma}^{(0)} -\frac{1}{2} \sum_{\bm{r},\bm{r}'\in\hat{\Gamma}} J(\bm{r},\bm{r}') \sigma(\bm{r}) \sigma(\bm{r}') \equiv H_{\Gamma}^{(0)} + V_{\Gamma},
\label{eq:H}
\eeq
where $H_{\Gamma}^{(0)}$ is the {\it reference Hamiltonian}, the condition on which will be specified later.
The second term of Eq.~(\ref{eq:H}) stands for the contribution of long-range interactions,
and the condition on the interaction potential $J(\bm{r},\bm{r}')$ will be also mentioned later.

In this paper, we mainly consider free boundary conditions, but the theorem presented in Sec.~\ref{sec:theorem} also holds for periodic boundary conditions as long as the distance $|\bm{r}-\bm{r}'|$ is interpreted by the minimum image convention (the distance between the two points appears in the crucial conditions~(\ref{eq:condition1}) and (\ref{eq:condition2})).

For convenience, we choose the zero point of energy so that, for any $\Gamma\subset\Gamma'$, $H_{\Gamma}=H_{\Gamma'}$ if $\sigma(\bm{r})=0$ for all $\bm{r}\in\Gamma'\backslash\Gamma$.
In other words, any spin in the ``null state'' $\sigma(\bm{r})=0$ does not contribute to the energy. 

The entropy $S(E,M,\Delta M;\Gamma)$ is defined as
\beq
S(E,M,\Delta M;\Gamma) = \ln \sum_{\bm{\sigma}_{\Gamma}\in\mathcal{S}^{N_{\Gamma}}} \theta(H_{\Gamma}\leq E) \theta\left( \sum_{\bm{r}\in\hat{\Gamma}}\sigma(\bm{r}) \in [M,M+\Delta M) \right).
\label{eq:S}
\eeq
The function $\theta$ is defined as
\beq
\theta(A)=
\begin{cases}
1 & \text{if $A$ is True,} \\
0 & \text{if $A$ is False.}
\end{cases}
\eeq
The magnetization is denoted by $M$, and the quantity $\Delta M$ is some number which is large enough to contain a large number of microscopic states with $\sum_{\bm{r}\in\hat{\Gamma}}\sigma(\bm{r}) \in [M,M+\Delta M)$,
but macroscopically very small.

Since the spin in the state $\sigma(\bm{r})=0$ does not contribute to the energy and the magnetization, for discrete spins we have
\beq
S(E,M,\Delta M,\Gamma)\leq S(E,M,\Delta M,\Gamma') 
\qquad \text{for any $\Gamma\subset\Gamma'$.}
\label{eq:entropy_increasing}
\eeq
This inequality is derived by restricting the spin configurations so that $\sigma(\bm{r})=0$ for all $\bm{r}\in\Gamma'\backslash\Gamma$.
In other words, all the allowed spin configurations on $\Gamma$ are included in those on $\Gamma'$,
and hence Eq.~(\ref{eq:entropy_increasing}) follows.
For continuous spins, the inequality~(\ref{eq:entropy_increasing}) does not hold as it is,
but a slightly modified inequality can be derived if we assume the continuity of the energy,
$|H_{\Gamma'}-H_{\Gamma}|\leq\epsilon\kappa N_{\Gamma'\backslash\Gamma}$
with some constant $\kappa>0$ if $|\sigma(\bm{r})|\leq\epsilon$ for all $\bm{r}\in\Gamma'\backslash\Gamma$.
The inequality in that case is given by
\beq
S(E+\kappa\epsilon N_{\Gamma'\backslash\Gamma},M,\Delta M-\epsilon N_{\Gamma'\backslash\Gamma};\Gamma) + N_{\Gamma'\backslash\Gamma} \ln\int_{-\epsilon}^{\epsilon}\eta(\sigma)d\sigma \leq S(E,M,\Delta M;\Gamma').
\label{eq:entropy_increasing2}
\eeq

The entropy density is given by
\beq
s(\varepsilon,m,\delta m;\Gamma) = \frac{1}{|\Gamma|} S(|\Gamma|\varepsilon,|\Gamma|m,|\Gamma|\delta m;\Gamma).
\label{eq:s}
\eeq

We consider the thermodynamic limit.
Now let us consider some fixed domain $\gamma\subset\mathbb{R}^d$ of unit volume, $|\gamma|=1$.
We set $\Gamma=L\gamma$, where the set $kA$ with $k\in\mathbb{R}$ and $A\subset\mathbb{R}^d$ is defined as $kA \equiv \{ \bm{x}\in\mathbb{R}^d : \bm{x}/k\in A\}$. Similarly, the set $A+\bm{a}$ with $A\subset\mathbb{R}^d$ and $\bm{a}\in\mathbb{R}^d$ is defined as $A+\bm{a} \equiv \{ \bm{x}\in\mathbb{R}^d : \bm{x}-\bm{a}\in A\}$.

By thermodynamic limit, we mean the limit of $L\rightarrow\infty$ with fixed values of $\varepsilon$ and $m$ and {\it with a fixed domain} $\gamma$.
It means that the system is made large with a fixed shape of the system.
Later we will see that in long-range interacting systems the thermodynamic limit depends on the shape of the system, $\gamma$.
As is well known, it is not the case in short-range interacting systems~\cite{Fermi_text,Ruelle_text}.
Thermodynamic limit of the entropy density is, if it exists, given by
\beq
s_{\gamma}(\varepsilon,m)=\lim_{\delta m\rightarrow 0}\lim_{L\rightarrow\infty} s(\varepsilon,m,\delta m;L\gamma),
\label{eq:s_TL}
\eeq
where $\delta m=\Delta M/N_{\Gamma}$, see Eq.~(\ref{eq:S}).
The aim of this paper is proving the existence of Eq.~(\ref{eq:s_TL}) and finding its simple expression.

Let us go back to our Hamiltonian, Eq.~(\ref{eq:H}), on which we impose some conditions.
The condition on $H_{\Gamma}^{(0)}$ is as follows.
Let $\Gamma=\Gamma_1\cup\Gamma_2$ with $\Gamma_1\cap\Gamma_2=\emptyset$ and define
\beq
H_{\Gamma_1,\Gamma_2}^{(0)}=H_{\Gamma}^{(0)}-H_{\Gamma_1}^{(0)}-H_{\Gamma_2}^{(0)},
\label{eq:ref_int}
\eeq
which expresses the interaction between subsystems $\Gamma_1$ and $\Gamma_2$.
Let us consider arbitrary two $d$-dimensional cubes of side $l$, $\Lambda_l^{(1)}$ and $\Lambda_l^{(2)}$ with $\Lambda_l^{(1)}\cap\Lambda_l^{(2)}=\emptyset$.
Then we assume that there exist positive constants $K>0$ and $\nu>0$ such that, for any such $d$-dimensional cubes,
\beq
\max_{\bm{\sigma}_{\Lambda_l^{(1)}},\bm{\sigma}_{\Lambda_l^{(2)}}} \left| H_{\Lambda_l^{(1)},\Lambda_l^{(2)}}^{(0)} \right| \leq \frac{Kl^{2d}}{R^{d+\nu}},
\label{eq:ref_condition}
\eeq
where $R$ is the distance between the center of $\Lambda_l^{(1)}$ and that of $\Lambda_l^{(2)}$, that is,
$$R=\frac{1}{l^d}\left|\int_{\Lambda_l^{(1)}}\bm{r}d^d\bm{r}-\int_{\Lambda_l^{(2)}}\bm{r}d^d\bm{r}\right|.$$
Intuitively, the above condition means that the reference Hamiltonian $H_{\Gamma}^{(0)}$ contains only short-range interactions.
We also assume that in the reference system the thermodynamic limit of the entropy density
\beq
s^{(0)}(\varepsilon,m)=\lim_{\delta m\rightarrow 0}\lim_{L\rightarrow\infty} s^{(0)}(\varepsilon,m,\delta m;L\gamma)
\label{eq:limit_reference_s}
\eeq
exists and is independent of $\gamma$.
This has been rigorously proven for a wide class of short-range interacting systems, see Ref.~\cite{Ruelle_text}.

It helps us to give a few examples of the reference Hamiltonian.
The Zeeman energy under the magnetic field is represented by $H_{\Gamma}^{(0)}=-h\sum_{\bm{r}\in\hat{\Gamma}}\sigma(\bm{r})$.
The Hamiltonian $H_{\Gamma}^{(0)}=-\kappa\sum_{\bm{r},\bm{r}'\in\hat{\Gamma}}\theta(|\bm{r}-\bm{r}'|=1)\sigma(\bm{r})\sigma(\bm{r}')$ stands for nearest-neighbor exchange interactions.

Next we mention the condition on $V_{\Gamma}$.
The potential $J(\bm{r},\bm{r}')$ represents long-range interactions between the spins at $\bm{r}$ and $\bm{r}'$.
By long-range interactions, we mean that $J(\bm{r},\bm{r}')$ is written in the following form,
\beq
J(\bm{r},\bm{r}')=\frac{1}{L^d}\phi\left(\frac{\bm{r}}{L},\frac{\bm{r}'}{L}\right)
\label{eq:long}
\eeq
for $\Gamma=L\gamma$ with $|\gamma|=1$.
The function $\phi$ is independent of $\Gamma$, symmetric $\phi(\bm{x},\bm{y})=\phi(\bm{y},\bm{x})$, and integrable on $\gamma\times\gamma$,
\beq
\int_{\gamma}d^d\bm{x}\int_{\gamma}d^d\bm{y}\phi(\bm{x},\bm{y}) = \mathcal{N}_{\phi,\gamma} < +\infty.
\label{eq:integrable}
\eeq
The value of $\mathcal{N}_{\phi,\gamma}$ is not important, so we put $\mathcal{N}_{\phi,\gamma}=1$.\footnote
{The sign of $\mathcal{N}_{\phi,\gamma}$ is important. By putting $\mathcal{N}_{\phi,\gamma}=1$, it is implicitly assumed that the interaction is ferromagnetic as a whole.}
Moreover, it is assumed that
\begin{align}
|\phi(\bm{x},\bm{y})| &\leq \frac{J}{|\bm{x}-\bm{y}|^{\alpha}},
\label{eq:condition1} \\
|\nabla_{\bm{x}}\phi(\bm{x},\bm{y})| &\leq \frac{J'}{|\bm{x}-\bm{y}|^{\alpha+1}}
\label{eq:condition2}
\end{align}
with some $J>0$, $J'>0$, and $\alpha\in[0,d)$.

When we consider the translationally invariant interaction potential, $\phi(\bm{x},\bm{y})=\phi(\bm{x}-\bm{y})$ and thus $J(\bm{r},\bm{r}')=L^{-d}\phi((\bm{r}-\bm{r}')/L)$.
It means that the interaction range and the size of the system are comparable.

The scaling form of Eq.~(\ref{eq:long}) makes the system extensive.
That is, a typical amount of energy due to long-range interactions is given by
$$-\frac{1}{2}\sum_{\bm{r},\bm{r}'\in\hat{\Gamma}}J(\bm{r},\bm{r}')\sigma(\bm{r})\sigma(\bm{r}')
\sim -L^{2d}\int_{\gamma}d^d\bm{x}\int_{\gamma}d^d\bm{y}\frac{1}{L^d}\phi(\bm{x},\bm{y})
\sim -L^d,
$$
which is of the order of the volume of the system.
For example, for the power-law interactions, $\phi(\bm{x},\bm{y})\propto |\bm{x}-\bm{y}|^{-\alpha}$ with $\alpha\in [0,d)$, $J(\bm{r},\bm{r}')$ has a scaling form of $J(\bm{r},\bm{r}')\propto L^{\alpha-d}|\bm{r}-\bm{r}'|^{-\alpha}$.
The factor $L^{\alpha-d}$ makes the interaction energy per spin finite when the interaction decays as $1/r^{\alpha}$.

The thermodynamic limit is an idealization to describe a real finite but large system.
The ideal limit should be taken in such a way that the thermodynamic properties of the system do not change by this limiting procedure.
In order to do that, the energy should be made extensive.
The procedure to make the system extensive by introducing the system-size dependence on $V_{\Gamma}$ as in Eq.~(\ref{eq:long}) is referred to as the ``Kac prescription''~\cite{Campa_review2009,Kac1963}.

In this paper, we only consider the microcanonical ensemble.
We can do it without loss of generality because if we can show that the microcanonical entropy has its thermodynamic limit,
it is automatically shown that the free energies in the canonical and the grandcanonical ensemble also have their thermodynamic limit. They are derived by the Legendre-Fenchel transformation from the microcanonical entropy.
On the other hand, it is pointed out that the inverse transformation, i.e. transformation from the canonical ensemble to the microcanonical ensemble, is impossible as a result of the ensemble inequivalence in long-range interacting systems~\cite{Campa_review2009,Les_Houches2009}.

Before presenting the main result, we briefly explain the additivity and its consequences.
The system is said to be additive if the following equality holds~\cite{Mori2014_quasi}:
\beq
s_{\gamma_1,\gamma_2}(\varepsilon,m_1,m_2) = \sup_{\substack{\varepsilon_1,\varepsilon_2: \\ \lambda\varepsilon_1+(1-\lambda)\varepsilon_2=\varepsilon}} \left[ \lambda s_{\gamma_1}(\varepsilon_1,m_1) + (1-\lambda) s_{\gamma_2}(\varepsilon_2,m_2) \right],
\label{eq:additivity}
\eeq
where $s_{\gamma_1,\gamma_2}(\varepsilon,m_1,m_2)$ is the entropy density of a state with the total energy density $\varepsilon$ and the magnetization densities $m_1$ and $m_2$ of the domains $\gamma_1$ and $\gamma_2$, respectively.

We can derive some important results from additivity, see Ref.~\cite{Mori2014_quasi} for the derivation.
Firstly, when the system is additive, the entropy density is independent of $\gamma$, the shape of the system:
\beq
s_{\gamma}(\varepsilon,m)=s_{\gamma'}(\varepsilon,m)\equiv s(\varepsilon,m).
\eeq
Secondly, we can show that the entropy density is a concave function of $\varepsilon$ and $m$:
\beq
s(\lambda\varepsilon_1+(1-\lambda)\varepsilon_2,\lambda m_1+(1-\lambda)m_2)
\geq \lambda s(\varepsilon_1,m_1) + (1-\lambda)s(\varepsilon_2,m_2).
\eeq

Concavity of the entropy ensures the ensemble equivalence, e.g., the microcanonical ensemble is equivalent to the canonical ensemble~\cite{Touchette_review2004}. 
As we have seen above, such important properties immediately follow from our definition of additivity.
Additivity in the sense of Eq.~(\ref{eq:additivity}) is, therefore, considered to be a fundamental property of macroscopic systems.

In short-range interacting systems with suitable conditions, it is rigorously shown that the system is additive~\cite{Ruelle_text}.
While it is not necessarily the case in long-range interacting systems~\cite{Campa_review2009,Les_Houches2009,Chavanis_review2006}.
As a result, in long-range interacting systems, the entropy density may depend on $\gamma$ and may not be concave.
A nonconcave entropy implies the ensemble inequivalence.

\section{Theorem on the thermodynamic limit}
\label{sec:theorem}

\subsection{Existence of the thermodynamic limit and the simple variational expression of the entropy density}

In this section we mention the theorem and discuss its consequence.
The theorem we now discuss is the following:
\begin{theorem}[Thermodynamic limit of the entropy density]
\label{theorem}
Consider the system described by Eqs.~(\ref{eq:H}) and (\ref{eq:long}) with the conditions given by Eqs.~(\ref{eq:condition1}) and (\ref{eq:condition2}).
Then the thermodynamic limit of the entropy density exists and is given by the following variational formula~\footnote
{The notation $\sup[A:B]$ means $\sup A$ under the condition $B$.}:
\begin{align}
s_{\gamma}(\varepsilon,m)=\sup_{\varepsilon(\cdot),m(\cdot)\in\mathcal{R}_{\gamma}} &\left[ \int_{\gamma}d^d\bm{x} s^{(0)}(\varepsilon(\bm{x}),m(\bm{x})) : 
\int_{\gamma}d^d\bm{x}m(\bm{x})=m,
\right. \nonumber \\
&\left.
-\frac{1}{2}\int_{\gamma}d^d\bm{x}\int_{\gamma}d^d\bm{y} \phi(\bm{x},\bm{y})m(\bm{x})m(\bm{y}) + \int_{\gamma}d^d\bm{x}\varepsilon(\bm{x})=\varepsilon
\right],
\label{eq:theorem}
\end{align}
where $s^{(0)}(\varepsilon,m)$ is the thermodynamic limit of the entropy density of the reference system described by $H_{\Gamma}^{(0)}$.
The set of Riemann integrable functions on $\gamma$ is denoted by $\mathcal{R}_{\gamma}$.
\end{theorem}

Equation~(\ref{eq:theorem}) means that thermodynamic properties can be described by the coarse-grained magnetization $m(\bm{x})$ and the coarse-grained energy density $\varepsilon(\bm{x})$.
In the proof of Theorem~\ref{theorem}, we will divide the original system into a large number of cells of side $l\ll L$.
We can show that the entropy density is almost unchanged by averaging out the spin variables within each cell (this averaging procedure is called the coarse graining).
This fact allows us to express the entropy density in the variational form as Eq.~(\ref{eq:theorem}).

We can give the explicit expression of $s^{(0)}(\varepsilon,m)$ for some simple cases.
When we consider the case $\mathcal{S}=\{0,1\}$, or $\sigma(\bm{r})=0$ or 1, and there is no short-range interactions, $H^{(0)}_{\Gamma}=0$, for example, we have
\beq
s^{(0)}(\varepsilon,m)=-m\ln m-(m+1)\ln(m+1) \quad \text{for } \varepsilon\geq 0
\eeq
and $s^{(0)}=0$ for $\varepsilon<0$.

In the canonical ensemble, it is rigorously proven that the free energy density defined by
\beq
f_{\gamma}(\beta,m) = \lim_{\delta m\rightarrow 0}\lim_{L\rightarrow\infty}
\left[ -\frac{1}{\beta}\ln \sum_{\bm{\sigma}_{L\gamma}\in\mathcal{S}^{N_{L\gamma}}} \theta\left( \frac{1}{L^d}\sum_{\bm{r}\in L\gamma\cap\mathbb{Z}^d}\sigma(\bm{r}) \in [m,m+\delta m) \right) e^{-\beta H_{L\gamma}}\right]
\eeq
is related to the entropy density via the Legendre-Fenchel transformation,
\beq
f_{\gamma}(\beta,m) = \inf_{\varepsilon}\left[ \varepsilon - \frac{1}{\beta}s_{\gamma}(\varepsilon,m) \right].
\label{eq:free_Legendre}
\eeq
By using Eq.~(\ref{eq:theorem}), Eq.~(\ref{eq:free_Legendre}) becomes
\beq
f_{\gamma}(\beta,m) = \inf_{m(\cdot)\in\mathcal{R}_{\gamma}} \left[ -\frac{1}{2}\int_{\gamma}d^d\bm{x} \int_{\gamma}d^d\bm{y} \phi(\bm{x},\bm{y})m(\bm{x})m(\bm{y}) + \int_{\gamma}d^d\bm{x} f^{(0)}(\beta, m(\bm{x}))\right],
\label{eq:theorem_free}
\eeq
where $f^{(0)}=\inf_{\varepsilon}[\varepsilon-s^{(0)}(\varepsilon,m)/\beta]$ is the free energy density of the reference system.
Equation (\ref{eq:theorem_free}) is the variational expression of the free energy density of a short- and long-range interacting spin system.

We can see Eqs.~(\ref{eq:theorem}) and (\ref{eq:theorem_free}) that the entropy density and the free energy density explicitly depend on $\gamma$, the shape of the system.
We have seen that in any additive system the entropy density is independent of $\gamma$ in the thermodynamic limit.
This fact, therefore, implies that a system with long-range interactions is in general not additive as expected.\footnote
{It is noted that the nonadditivity does not imply the shape dependence of the entropy.
In infinite-range models, the spacial geometry is not important and the entropy density does not depend on $\gamma$, but they are nonadditive.
The shape-dependent entropy density always implies nonadditivity.}

\section{Proof of the Theorem}
\label{sec:proof}

\subsection{Outline}

In this section we give a proof of Theorem~\ref{theorem}.
In long-range interacting systems, it is expected that short length-scale structure is not essential for thermodynamic properties.
Hence the method of {\it coarse graining} is a powerful tool to examine macroscopic properties of long-range interacting systems~\cite{Lebowitz-Penrose1966,van_Kampen1964,Barre2005}.
First we show that the procedure of coarse graining is justified and then show that the entropy density calculated by the coarse graining has a limiting value predicted by Theorem~\ref{theorem} in the thermodynamic limit.

\subsection{Two lemmas and the proof of the Theorem}

\begin{figure}[t]
  \begin{center}
    \includegraphics[clip, width=9cm]{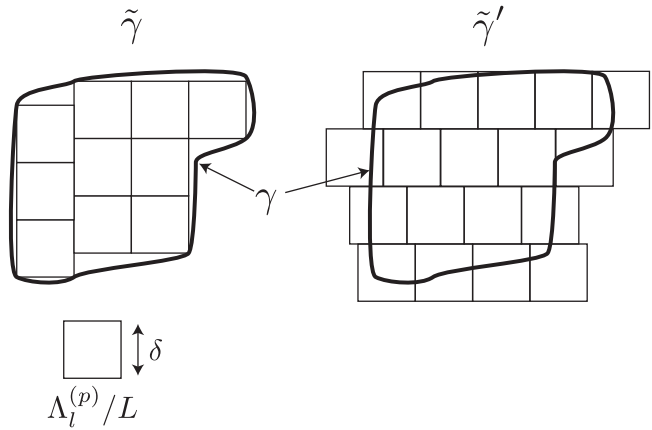}
    \caption{Schematic pictures of the domains $\tilde{\gamma}$ (left) and $\tilde{\gamma}'$ (right) for a given two-dimensional domain $\gamma$ with a unit volume (the region inside of the thick lines).
Each square of side $\delta$ expresses $\Lambda_l^{(p)}/L$.}
    \label{fig:domains}
  \end{center}
\end{figure}

We approximate $\Gamma$ by an ensemble of $d$-dimensional cubes of side $l$, each of which is denoted by $\Lambda_l^{(p)}$, $p=1,2,\dots$, with $\Lambda_l^{(p)}\cap\Lambda_l^{(q)}=\emptyset$.
We consider the two ways of approximations, see Fig.~\ref{fig:domains}.
Firstly, we fill $\Gamma$ with $\Lambda_l^{(p)}$ so that $\tilde{\Gamma}=\cup_p\Lambda_l^{(p)}\subset\Gamma$ has the maximum volume.
Secondly, we consider $\tilde{\Gamma}'=\cup_p\Lambda_l^{(p)}$ with the least volume satisfying $\tilde{\Gamma}'\supset\Gamma$.
We use the same symbol but $\Lambda_l^{(p)}$ identifying $\tilde{\Gamma}$ and $\Lambda_l^{(p)}$ identifying $\tilde{\Gamma}'$ may be different.
The domains $\tilde{\gamma}$ and $\tilde{\gamma}'$ are defined by $\tilde{\Gamma}=L\tilde{\gamma}$ and $\tilde{\Gamma}'=L\tilde{\gamma}'$.
Of course, $\tilde{\gamma}\subset\gamma\subset\tilde{\gamma}'$, where $\gamma$ is defined by $\Gamma=L\gamma$.
The domains $\tilde{\gamma}$ and $\tilde{\gamma}'$ are ones that approximate $\gamma$ by an ensemble of $d$-dimensional cubes of side $\delta=l/L$.
We assume that $\tilde{\gamma},\tilde{\gamma}'\rightarrow\gamma$ in the limit of $\delta\rightarrow +0$.

The coarse-grained Hamiltonian is obtained by averaging out $\sigma(\bm{r})$ within each cell $\Lambda_l^{(p)}$:
\beq
H_{\tilde{\gamma}}^{(\delta,l)}=H_{\tilde{\Gamma}}^{(0)} - \frac{1}{2L^d}\sum_{\substack{p,q \\ (\Lambda_l^{(p)},\Lambda_l^{(q)}\subset\tilde{\Gamma})}}\phi^{(\delta,l)}_{pq}M_pM_q,
\label{eq:cg_H}
\eeq
where
\beq
\phi_{pq}^{(\delta,l)}=\frac{1}{l^d}\sum_{\bm{r}\in\hat{\Lambda}_l^{(p)}} \frac{1}{l^d}\sum_{\bm{r}'\in\hat{\Lambda}_l^{(q)}}\phi\left(\frac{\bm{r}}{L},\frac{\bm{r}'}{L}\right)
\eeq
and
\beq
M_p=l^dm_p=\sum_{\bm{r}\in\hat{\Lambda}_l^{(p)}}\sigma(\bm{r}).
\eeq
Here, $\hat{\Lambda}_l^{(p)}=\Lambda_l^{(p)}\cap\mathbb{Z}^d$.
The coarse-grained Hamiltonian on $\tilde{\gamma}'$ is obtained by replacing $\tilde{\gamma}$ by $\tilde{\gamma}'$.

Since $L=l/\delta$, the difference between the exact Hamiltonian and the coarse-grained one,
\beq
\frac{1}{L^d}\max_{\bm{\sigma}_{\tilde{\Gamma}}\in\mathcal{S}^{N_{\tilde{\Gamma}}}} \left| H_{\tilde{\Gamma}}-H_{\tilde{\gamma}}^{(\delta,l)} \right| \equiv \Delta_{\tilde{\gamma}}^{(\delta,l)}
\eeq
are determined by $\delta$ and $l$.
If $\Delta_{\tilde{\gamma}}^{(\delta,l)}$ and $\Delta_{\tilde{\gamma}'}^{(\delta,l)}$ can be made vanishingly small in the thermodynamic limit, the procedure of coarse graining is justified.
Indeed we can show the following lemma, whose proof is given in Sec.~\ref{sec:Lemma1},

\begin{lemma}[Justification of the coarse graining]
For any given $\gamma$, there exists $\Delta_{\delta}>0$ depending on $\delta$ such that $\lim_{\delta\rightarrow 0}\Delta_{\delta}=0$ and $\Delta_{\tilde{\gamma}}^{(\delta,l)}, \Delta_{\tilde{\gamma}'}^{(\delta,l)}\leq\Delta_{\delta}$ for all $l>0$.
\label{lemma1}  
\end{lemma}

This lemma tells us that
\beq
\lim_{\delta\rightarrow 0}\frac{1}{L^d} \max_{\bm{\sigma}_{\tilde{\Gamma}}\in\mathcal{S}^{N_{\tilde{\Gamma}}}} \left| H_{L\tilde{\gamma}}-H_{\tilde{\gamma}}^{(\delta,l)} \right| = 0,
\eeq
where $L=l/\delta$, and its convergence is uniform with respect to $l$.

From Eq.~(\ref{eq:entropy_increasing}), we have
\beq
S(E,M,\Delta M;\tilde{\Gamma}) \leq S(E,M,\Delta M;\Gamma) \leq S(E,M,\Delta M;\tilde{\Gamma}')
\eeq
for discrete spins.
For continuous spins, the corresponding inequality is obtained by using Eq.~(\ref{eq:entropy_increasing2}), and it is slightly different from the above one. 
However, we can show the theorem by following the same line of the proof for discrete spins and finally taking the limit of $\epsilon\rightarrow +0$ (Remember that $\epsilon$ appears in Eq.~(\ref{eq:entropy_increasing2})).
Therefore, hereafter we focus on the case of discrete spins for simplicity.

By using Lemma~\ref{lemma1}, we obtain
\beq
S^{(\delta,l)}(E-L^d\Delta_{\delta},M,\Delta M;\tilde{\Gamma})
\leq S(E,M,\Delta M;\Gamma)
\leq S^{(\delta,l)}(E+L^d\Delta_{\delta},M,\Delta M;\tilde{\Gamma}').
\label{eq:inequality_S}
\eeq
Here, $S^{(\delta,l)}$ is the entropy calculated by $H^{(\delta,l)}$.
The corresponding entropy density is denoted by $s^{(\delta,l)}(\varepsilon,m,\delta m;\tilde{\Gamma})$.
In terms of the entropy densities, the inequality~(\ref{eq:inequality_S}) becomes
\beq
s^{(\delta,l)}(\varepsilon-\Delta_{\delta},\tilde{m},\delta\tilde{m};L\tilde{\gamma})
\leq s(\varepsilon,m,\delta m;L\gamma)
\leq s^{(\delta,l)}(\varepsilon+\Delta_{\delta},\tilde{m}',\delta\tilde{m}';L\tilde{\gamma}'),
\label{eq:inequality_s}
\eeq
where $\tilde{m}=M/|\tilde{\Gamma}|=m/|\tilde{\gamma}|$, $\delta\tilde{m}=\delta m/|\tilde{\gamma}|$, $\tilde{m}'=m/|\tilde{\gamma}'|$, and $\delta\tilde{m}'=\delta m/|\tilde{\gamma}'|$.

We take the limit of $\delta\rightarrow 0$ after $l\rightarrow\infty$ is taken.
In this limit, $(\tilde{\gamma},\tilde{\gamma}')\rightarrow\gamma$, $\Delta_{\delta}\rightarrow 0$, $(\tilde{m},\tilde{m}')\rightarrow m$, $(\delta\tilde{m},\delta\tilde{m}')\rightarrow\delta m$.
Thus if
\beq
\tilde{s}_{\gamma}(\varepsilon,m)=\lim_{\delta m\rightarrow 0}\lim_{\delta\rightarrow 0}\lim_{l\rightarrow\infty}s^{(\delta,l)}(\varepsilon,m,\delta m;\tilde{\Gamma})
\label{eq:cg_s}
\eeq
exists, the thermodynamic limit of the entropy density also exists and $s_{\gamma}(\varepsilon,m)=\tilde{s}_{\gamma}(\varepsilon,m)$ from the inequality~(\ref{eq:inequality_s}).
In Sec.~\ref{sec:Lemma2} we show this fact summarized in the following lemma.

\begin{lemma}[Existence of the thermodynamic limit of the coarse-grained entropy density]
\label{lemma2}
The limit of Eq.~(\ref{eq:cg_s}) exists and is expressed as
\begin{align}
\tilde{s}_{\gamma}(\varepsilon,m) = \sup_{\varepsilon(\cdot),m(\cdot)\in\mathcal{R}_{\gamma}}
\left[ \int_{\gamma}d^d\bm{x}s^{(0)}(\varepsilon(\bm{x}),m(\bm{x})) : \int_{\gamma}d^d\bm{x}m(\bm{x})=m, \right. \\
\left. \int_{\gamma}d^d\bm{x}\varepsilon(\bm{x})
-\frac{1}{2}\int_{\gamma}d^d\bm{x}\int_{\gamma}d^d\bm{y}\phi(\bm{x},\bm{y})m(\bm{x})m(\bm{y})
=\varepsilon \right].
\end{align}
\end{lemma}

By combining Lemma~\ref{lemma1} and Lemma~\ref{lemma2}, we obtain Theorem~\ref{theorem}.

\subsection{Proof of Lemma 1}
\label{sec:Lemma1}

\begin{figure}[t]
\begin{center}
\includegraphics[clip,width=6cm]{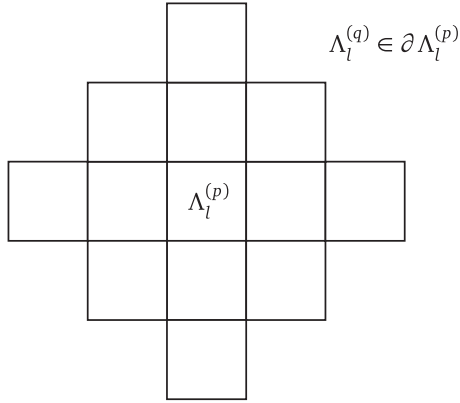}
\caption{The set of $\Lambda_l^{(q)}\in\d\Lambda_l^{(p)}$ in the case that the cells are tightly arranged on the two-dimensional space.
The central cell is $\Lambda_l^{(p)}$ and $\Lambda_l^{(p)}$ itself is also the element of $\d\Lambda_l^{(p)}$.}
\label{fig:del_Lambda}
\end{center}
\end{figure}

We evaluate the upper bound of $\Delta^{(\delta,l)}_{\tilde{\gamma}}$, which is given by
\begin{align}
&\Delta^{(\delta,l)}_{\tilde{\gamma}}=\frac{1}{L^d}\max_{\bm{\sigma}_{\tilde{\Gamma}}\in\mathcal{S}^{N_{\tilde{\Gamma}}}}\left|H_{\tilde{\Gamma}}-H^{(\delta,l)}_{\tilde{\gamma}}\right|
\nonumber \\
&=\frac{1}{L^d}\max_{\bm{\sigma}_{\tilde{\Gamma}}\in\mathcal{S}^{N_{\tilde{\Gamma}}}}
\left|\frac{1}{2L^{2d}}\sum_{\substack{p,q \\ (\Lambda_l^{(p)},\Lambda_l^{(q)}\subset\tilde{\Gamma})}} \sum_{\substack{\bm{r}_1\in\hat{\Lambda}_l^{(p)} \\ \bm{r}_2\in\hat{\Lambda}_l^{(q)}}}\sigma(\bm{r}_1)\sigma(\bm{r}_2)
\sum_{\substack{\bm{r}_3\in\hat{\Lambda}_l^{(p)} \\ \bm{r}_4\in\hat{\Lambda}_l^{(q)}}} \frac{1}{l^{2d}}
\left[ \phi\left(\frac{\bm{r}_1}{L},\frac{\bm{r}_2}{L}\right) - \phi\left(\frac{\bm{r}_3}{L},\frac{\bm{r}_4}{L} \right) \right] \right|
\nonumber \\
&\leq \frac{\sigma_{\rm max}^2}{2L^{2d}l^{2d}}\sum_{p,q}\sum_{\substack{\bm{r}_1,\bm{r}_3\in\hat{\Lambda}_l^{(p)} \\ \bm{r}_2,\bm{r}_4\in\hat{\Lambda}_l^{(q)}}}
\left| \phi\left(\frac{\bm{r}_1}{L},\frac{\bm{r}_2}{L}\right) - \phi\left(\frac{\bm{r}_3}{L},\frac{\bm{r}_4}{L} \right) \right|.
\end{align}
We divide the summation over $q$ into that with $\Lambda_l^{(q)}\in\d\Lambda_l^{(p)}$ and that with $\Lambda_l^{(q)}\notin\d\Lambda_l^{(p)}$, where $\d\Lambda_l^{(p)}$ is defined as
\beq
\Lambda_l^{(q)}\in\d\Lambda_l^{(p)} \Leftrightarrow r_{pq}<2\sqrt{d}l,
\eeq
see Fig.~\ref{fig:del_Lambda} for visualizing the set $\{\Lambda_l^{(q)}\in\d\Lambda_l^{(p)}\}$ in the case that the cells are tightly arranged on the two-dimensional space.
The distance between the central points of $\Lambda_l^{(p)}$ and $\Lambda_l^{(q)}$ has been denoted by $r_{pq}$.
Then we have
\begin{align}
\Delta^{(\delta,l)}_{\tilde{\gamma}}\leq&
\frac{\sigma_{\rm max}^2}{2L^{2d}l^{2d}}\sum_{\substack{p,q \\ \Lambda_l^{(q)}\in\d\Lambda_l^{(p)}}}\sum_{\substack{\bm{r}_1,\bm{r}_3\in\hat{\Lambda}_l^{(p)} \\ \bm{r}_2,\bm{r}_4\in\hat{\Lambda}_l^{(q)}}}
\left| \phi\left(\frac{\bm{r}_1}{L},\frac{\bm{r}_2}{L}\right) - \phi\left(\frac{\bm{r}_3}{L},\frac{\bm{r}_4}{L} \right) \right|
\nonumber \\
&+\frac{\sigma_{\rm max}^2}{2L^{2d}l^{2d}}\sum_{\substack{p,q \\ \Lambda_l^{(q)}\notin\d\Lambda_l^{(p)}}}\sum_{\substack{\bm{r}_1,\bm{r}_3\in\hat{\Lambda}_l^{(p)} \\ \bm{r}_2,\bm{r}_4\in\hat{\Lambda}_l^{(q)}}}
\left| \phi\left(\frac{\bm{r}_1}{L},\frac{\bm{r}_2}{L}\right) - \phi\left(\frac{\bm{r}_3}{L},\frac{\bm{r}_4}{L} \right) \right|
\nonumber \\
\equiv& \Delta_1+\Delta_2
\end{align}

Let us first evaluate $\Delta_1$.
From Eq.~(\ref{eq:condition1}),
\begin{align}
\left| \phi\left(\frac{\bm{r}_1}{L},\frac{\bm{r}_2}{L}\right) - \phi\left(\frac{\bm{r}_3}{L},\frac{\bm{r}_4}{L} \right) \right|
&\leq \left| \phi\left(\frac{\bm{r}_1}{L},\frac{\bm{r}_2}{L}\right)\right| + \left| \phi\left(\frac{\bm{r}_3}{L},\frac{\bm{r}_4}{L} \right) \right|
\nonumber \\
&\leq JL^{\alpha}\left(\frac{1}{|\bm{r}_1-\bm{r}_2|^{\alpha}}+\frac{1}{|\bm{r}_3-\bm{r}_4|^{\alpha}}\right).
\end{align}
By substituting it into the expression of $\Delta_1$, we obtain
\beq
\Delta_1 \leq \frac{\sigma_{\rm max}^2}{L^{2d}}\sum_p\sum_{\substack{q \\ (\Lambda_l^{(q)}\in\d\Lambda_l^{(p)})}} \sum_{\bm{r}_1\in\hat{\Lambda}_l^{(p)}} \sum_{\substack{\bm{r}_2\in\hat{\Lambda}_l^{(q)} \\ (\bm{r}_1\neq\bm{r}_2)}} \frac{JL^{\alpha}}{|\bm{r}_1-\bm{r}_2|^{\alpha}}.
\eeq
For $\bm{r}_1\in\Lambda_l^{(p)}$ and $\bm{r}_2\in\Lambda_l^{(q)}$ with $\Lambda_l^{(q)} \in \d\Lambda_l^{(p)}$, $|\bm{r}_1-\bm{r}_2|\leq \sqrt{d}l+r_{pq}<3\sqrt{d}l$.
Hence,
\begin{align}
\Delta_1&\leq\frac{\sigma_{\rm max}^2}{L^{2d}} \sum_{\substack{\bm{r}_1,\bm{r}_2\in\hat{\tilde{\Gamma}} \\ (|\bm{r}_1-\bm{r}_2|<3\sqrt{d}l)}} \frac{JL^{\alpha}}{|\bm{r}_1-\bm{r}_2|^{\alpha}}
\nonumber \\
&\leq \frac{\sigma_{\rm max}^2J}{L^{2d-\alpha}}\left(\sum_{\bm{r}_1\in\hat{\tilde{\Gamma}}}1\right) \int_0^{3\sqrt{d}l}dr S_dr^{d-\alpha-1}
\nonumber \\
&= \frac{(3\sqrt{d})^{d-\alpha}\sigma_{\rm max}^2JS_d}{d-\alpha} \frac{N_{\tilde{\Gamma}}}{L^d} \delta^{d-\alpha}
\nonumber \\
&\approx \frac{(3\sqrt{d})^{d-\alpha}\sigma_{\rm max}^2JS_d}{d-\alpha}|\tilde{\gamma}| \delta^{d-\alpha},
\label{eq:Delta1}
\end{align}
where $S_d=2\pi^{(d+1)/2}/\Gamma((d+1)/2)$ is the surface area of the $d$-dimensional unit cube.
This upper limit is independent of $l$ and going to zero in the limit of $\delta\rightarrow 0$.

Next, we shall evaluate $\Delta_2$.
By the mean-value theorem, there exists $u\in[0,1]$ such that
\beq
\phi\left(\frac{\bm{r}_1-\bm{r}_2}{L}\right)-\phi\left(\frac{\bm{r}_3-\bm{r}_4}{L}\right) = \left[\nabla_{\bm{x}}\phi(\bm{x},\bm{y})\right]\cdot\frac{\bm{r}_1-\bm{r}_3}{L} + \left[\nabla_{\bm{y}}\phi(\bm{x},\bm{y})\right]\cdot\frac{\bm{r}_2-\bm{r}_4}{L},
\label{eq:mean_value}
\eeq
with $\bm{x}=[(1-u)\bm{r}_1+u\bm{r}_3]/L$ and $\bm{y}=[(1-u)\bm{r}_2+u\bm{r}_4]/L$.
Because $\bm{r}_1,\bm{r}_3\in\Lambda_l^{(p)}$ and $\bm{r}_2,\bm{r}_4\in\Lambda_l^{(q)}$, $L\bm{x}\in\Lambda_l^{(p)}$ and $L\bm{y}\in\Lambda_l^{(q)}$.
Here, by the triangle inequality, $L|\bm{x}-\bm{y}|\geq r_{pq}-\sqrt{d}l$.
Since $\Lambda_l^{(q)}\notin\d\Lambda_l^{(p)}$, $r_{pq}\geq 2\sqrt{d}l$ and thus $|\bm{x}-\bm{y}|\geq r_{pq}/2L$.
Moreover, $|\bm{r}_1-\bm{r}_3|\leq\sqrt{d}l$ and $|\bm{r}_2-\bm{r}_4|\leq\sqrt{d}l$.
Due to the condition (\ref{eq:condition2}) and Eq.~(\ref{eq:mean_value}),
\begin{align}
\left|\phi\left(\frac{\bm{r}_1-\bm{r}_2}{L}\right)-\phi\left(\frac{\bm{r}_3-\bm{r}_4}{L}\right)\right| &\leq \frac{J'}{|\bm{x}-\bm{y}|^{\alpha+1}}\left(\frac{|\bm{r}_1-\bm{r}_3|}{L}+\frac{|\bm{r}_2-\bm{r}_4|}{L}\right)
\nonumber \\
&\leq 2^{\alpha+2}\sqrt{d}J'\left(\frac{L}{r_{pq}}\right)^{\alpha+1}\delta.
\end{align}
Because $\Lambda_l^{(q)}\notin\d\Lambda_l^{(p)}$, $r_{pq}\geq 2\sqrt{d}l$, and hence $r_{pq}/L\geq 2\sqrt{d}\delta$.
Thus we can evaluate $\Delta_2$ as
\begin{align}
\Delta_2 &\leq 2^{\alpha+1}\sqrt{d}J'\sigma_{\rm max}^2\delta^{2d+1}\sum_p\sum_{\substack{q \\ (\Lambda_l^{(q)}\notin\d\Lambda_l^{(p)})}} \left(\frac{L}{r_{pq}}\right)^{\alpha+1}
\nonumber \\
&\leq 2^{\alpha+1}\sqrt{d}J'\sigma_{\rm max}^2\delta^{d+1}\sum_p\int_{2\sqrt{d}\delta}^{x_{\rm max}}dx S_d x^{d-\alpha-2}
\nonumber \\
&=A|\tilde{\gamma}|\delta\int_{2\sqrt{d}\delta}^{x_{\rm max}}dx x^{d-\alpha-2}
\end{align}
where $x_{\rm max}$ is defined as $x_{\rm max}=\max_{\bm{x},\bm{y}\in\tilde{\gamma}}|\bm{x}-\bm{y}|$, which is assumed to be finite,
and $A=2^{\alpha+1}\sqrt{d}J'\sigma_{\rm max}^2S_d$.
By evaluating the integral, we obtain
\begin{subnumcases}{\Delta_2\leq}
A\dfrac{x_{\rm max}^{d-\alpha-1}}{d-\alpha-1}|\tilde{\gamma}|\delta & ($\alpha< d-1$), \label{eq:Delta2a}\\
A|\tilde{\gamma}|\delta\ln\dfrac{x_{\rm max}}{2\sqrt{d}\delta} & ($\alpha=d-1$), \label{eq:Delta2b}\\
A\dfrac{1}{(\alpha-d+1)(2\sqrt{d})^{\alpha-d+1}}|\tilde{\gamma}|\delta^{d-\alpha} & ($d-1<\alpha<d$), \label{eq:Delta2c}
\end{subnumcases}
In any case, as long as $\alpha<d$, $\Delta_2\rightarrow 0$ in the limit of $\delta\rightarrow 0$.
The convergence is uniform with respect to $l$ because the derived upper bound is independent of $l$.

By collecting the results for $\Delta_1$ and $\Delta_2$, we complete the proof of Lemma~\ref{lemma1}.

\subsection{Proof of Lemma 2}
\label{sec:Lemma2}

We prove Lemma~\ref{lemma2} by evaluating the upper bound and the lower bound of $s^{(\delta,l)}(\varepsilon,m,\delta m;\tilde{\Gamma})$ and showing that these bounds become indistinguishable in a suitable limit.

We decompose the reference Hamiltonian as
\beq
H^{(0)}_{\tilde{\Gamma}}=\sum_{\substack{p \\ (\Lambda_l^{(p)}\subset\tilde{\Gamma})}} H_{pp}^{(0)}(\bm{\sigma}_{\Lambda_l^{(p)}})+\sum_{\substack{p<q \\ (\Lambda_l^{(p)},\Lambda_l^{(q)}\subset\tilde{\Gamma})}}H_{pq}^{(0)}(\bm{\sigma}_{\Lambda_l^{(p)}},\bm{\sigma}_{\Lambda_l^{(q)}}),
\eeq
and we write $H^{(\delta,l)}_{\tilde{\gamma}}=H_{\tilde{\Gamma}}^{(0)}+V^{(\delta,l)}_{\tilde{\gamma}}$.
Because of the condition (\ref{eq:ref_condition}), there are some $K>0$ and $\nu>0$ such that
\beq
|H_{pq}^{(0)}|\leq\frac{Kl^{2d}}{r_{pq}^{d+\nu}} \equiv E_{pq}^{(0)}
\eeq
with
\beq
r_{pq}=\frac{1}{l^d} \left| \int_{\Lambda_l^{(p)}}\bm{r}d^d\bm{r}-\int_{\Lambda_l^{(q)}}\bm{r}d^d\bm{r} \right|.
\eeq
The coarse-grained entropy density is explicitly given by
\beq
s^{(\delta,l)}(\varepsilon,m,\delta m;\tilde{\Gamma})=\frac{1}{|\tilde{\Gamma}|} \ln \sum_{\bm{\sigma}_{\tilde{\Gamma}}\in\mathcal{S}^{N_{\tilde{\Gamma}}}} \theta\left(H^{(\delta,l)}_{\tilde{\gamma}}\leq|\tilde{\Gamma}|\varepsilon\right) \theta\left(\frac{1}{|\tilde{\Gamma}|}\sum_{\bm{r}\in\tilde{\Gamma}}\sigma(\bm{r})\in [m,m+\delta m)\right).
\label{eq:cg_s2}
\eeq

We have
\begin{align}
\theta\left(\sum_pH_{pp}^{(0)}+V^{(\delta,l)}_{\tilde{\gamma}}\leq|\tilde{\Gamma}|\varepsilon -\sum_{p<q}E_{pq}^{(0)}\right)
\leq \theta\left(H^{(\delta,l)}_{\tilde{\gamma}}\leq|\tilde{\Gamma}|\varepsilon\right)
\nonumber \\
\leq \theta\left(\sum_pH_{pp}^{(0)}+V^{(\delta,l)}_{\tilde{\gamma}}\leq|\tilde{\Gamma}|\varepsilon +\sum_{p<q}E_{pq}^{(0)}\right).
\label{eq:ineq_H}
\end{align}
We define $\varepsilon_g^{(0)}>-\infty$ as the possible minimum value of $H_{pp}^{(0)}/l^d$.
Decompose the possible values of energy and magnetization as
\beq
\frac{1}{l^d}H_{pp}^{(0)}-\varepsilon_g^{(0)}\in [n_p\delta\varepsilon,(n_p+1)\varepsilon)
\eeq
and
\beq
\frac{1}{l^d}\sum_{\bm{r}\in\Lambda_l^{(p)}}\sigma(\bm{r})=m_p\in\left[k_p\frac{\delta m}{2},(k_p+1)\frac{\delta m}{2}\right)
\eeq
with integers $\{ n_p\}$ and $\{ k_p\}$.
An arbitrary positive constant $\delta\varepsilon$ has been introduced.

Because $|\sigma(\bm{r})|\leq\sigma_{\rm max}$ or $|m_p|\leq \sigma_{\rm max}$, we can restrict the summation over $k_p$ to $k_{\rm min}\leq k_p \leq k_{\rm max}$.
Here, $k_{\rm min}$ is the maximum integer satisfying $k_{\rm min}\leq -2\sigma_{\rm max}/\delta m$ and $k_{\rm max}$ is the minimum integer satisfying $k_{\rm max}>2\sigma_{\rm max}/\delta m -1$.

We can also restrict the summation over $n_p$.
Since $H_{pp}^{(0)}-l^d\varepsilon_g^{(0)}\geq 0$, we have $n_p\geq 0$.
The inequality appearing in Eq.~(\ref{eq:ineq_H}),
$$\sum_pH_{pp}^{(0)}+V^{(\delta,l)}_{\tilde{\gamma}}\leq|\tilde{\Gamma}|\varepsilon -\sum_{p<q}E_{pq}^{(0)}$$
leads to
\beq
\sum_pn_p\leq\frac{|\tilde{\Gamma}|(\varepsilon-\varepsilon_g^{(0)}-\delta\varepsilon) + \left|V^{(\delta,l)}_{\tilde{\gamma}}\right| + \sum_{p<q}E_{pq}^{(0)}}{l^d\delta\varepsilon}.
\label{eq:bound_n}
\eeq
Here,
\begin{align}
\left|V^{(\delta,l)}_{\tilde{\gamma}}\right|
&= \left|-\frac{1}{2L^d}\sum_{pq}\phi^{(\delta,l)}_{p,q}M_pM_q\right|
\nonumber \\
&\leq \frac{L^d\sigma_{\rm max}^2}{2}\frac{1}{L^{2d}}\sum_{p,q}\left|\phi_{pq}^{(\delta,l)}\right|
\nonumber \\
&\leq \frac{L^d\sigma_{\rm max}^2}{2}\frac{1}{L^{2d}}\sum_{p,q}\sum_{\bm{r}\in\Lambda_l^{(p)}} \sum_{\bm{r}'\in\Lambda_l^{(q)}}\left|\phi\left(\frac{\bm{r}}{L},\frac{\bm{r}'}{L}\right)\right|
\nonumber \\
&\leq \frac{L^d\sigma_{\rm max}^2}{2}\frac{1}{L^{2d}}\sum_{\bm{r},\bm{r}'\in\hat{\tilde{\Gamma}}} \frac{J}{\left|\frac{\bm{r}-\bm{r}'}{L}\right|^{\alpha}}
\nonumber \\
&\leq \frac{L^d\sigma_{\rm max}^2}{2}\int_{\tilde{\gamma}}d^d\bm{x}\int_{\tilde{\gamma}}d^d\bm{y} \frac{J}{|\bm{x}-\bm{y}|^{\alpha}}
\equiv \frac{L^d\sigma_{\rm max}^2}{2}v,
\label{eq:bound_V}
\end{align}
where $v$ is a constant independent of $\delta$ and $l$.
We have used Eq.~(\ref{eq:condition1}).
As for $E_{pq}^{(0)}$, we can evaluate as
\begin{align}
\sum_{p<q}E_{pq}^{(0)}
&= Kl^{2d}\sum_{p<q}\frac{1}{r_{pq}^{d+\nu}}
\nonumber \\
&= \frac{K}{2}l^{2d}\sum_p\sum_{\substack{q \\ (\Lambda_l^{(q)}\in\d\Lambda_l^{(p)})}}\frac{1}{r_{pq}^{d+\nu}} + \frac{K}{2}l^{2d}\sum_p\sum_{\substack{q \\ (\Lambda_l^{(q)}\notin\d\Lambda_l^{(p)})}}\frac{1}{r_{pq}^{d+\nu}}
\nonumber \\
&\leq \frac{K}{2}l^{2d}\sum_p\sum_{\substack{q \\ (\Lambda_l^{(q)}\in\d\Lambda_l^{(p)})}}\frac{1}{l^{d+nu}} + \frac{K}{2}l^d\sum_p\int_{2\sqrt{d}l}^{\infty}dx S_dx^{d-1}\frac{1}{x^{d+\nu}}
\nonumber \\
&\leq |\tilde{\Gamma}|\frac{K}{2}\left[n_d+\frac{S_d}{\nu (2\sqrt{d})^{\nu}}\right]l^{-\nu}.
\label{eq:bound_E}
\end{align}
Here, $n_d$ is the maximum number of elements of $\d\Lambda_l^{(p)}$, that is, $n_d=\max_p \sum_{q (\Lambda_l^{(q)}\in\d\Lambda_l^{(p)})}1$.
By using Eqs.~(\ref{eq:bound_V}) and (\ref{eq:bound_E}), Eq.~(\ref{eq:bound_n}) becomes
\begin{align}
\sum_pn_p
& \leq\frac{|\tilde{\Gamma}|}{l^d\delta\varepsilon}\left\{\varepsilon-\varepsilon_g^{(0)}-\delta\varepsilon +\frac{L^d}{2|\tilde{\Gamma}|}\sigma_{\rm max}^2v +\frac{K}{2}\left[n_d+\frac{S_d}{\nu (2\sqrt{d})^{\nu}}\right]l^{-\nu}\right\}
\nonumber \\
&\equiv N_{\rm sum}.
\label{eq:Nsum}
\end{align}
Since $|\tilde{\Gamma}|\approx L^d$ for small values of $\delta=l/L$,
$N_{\rm sum}$ is a quantity of order $\delta^{-d}$.
From Eq.~(\ref{eq:Nsum}), by remembering that the number of $p$ with $\Lambda_l^{(p)}\subset\tilde{\Gamma}$ is $|\tilde{\Gamma}|/l^d$ and $n_p\geq 0$, we obtain
\beq
\sum_{\{n_p\}} \theta\left(\sum_pH_{pp}^{(0)}+V^{(\delta,l)}_{\tilde{\gamma}}\leq|\tilde{\Gamma}|\varepsilon +\sum_{p<q}E_{pq}^{(0)}\right)
\leq \frac{(N_{\rm sum}+|\tilde{\Gamma}|/l^d)!}{N_{\rm sum}!(|\tilde{\Gamma}|/l^d)!},
\eeq
or
\begin{align}
&\ln\sum_{\{n_p\}}\theta\left(\sum_pH_{pp}^{(0)}+V^{(\delta,l)}_{\tilde{\gamma}}\leq|\tilde{\Gamma}|\varepsilon +\sum_{p<q}E_{pq}^{(0)}\right)
\nonumber \\
&\leq \left(N_{\rm sum}+\frac{|\tilde{\Gamma}|}{l^d}+1\right)\ln\left(N_{\rm sum}+\frac{|\tilde{\Gamma}|}{l^d}+1\right) - N_{\rm sum}\ln N_{\rm sum} - \frac{|\tilde{\Gamma}|}{l^d}\ln\frac{|\tilde{\Gamma}|}{l^d}.
\end{align}
We have used the inequality $n\ln n-n\leq \ln n!\leq (n+1)\ln(n+1)-n$.
By putting $N_{\rm sum}=\delta^{-d}n_{\rm sum}$ and $|\tilde{\Gamma}|/l^d=\delta^{-d}\tilde{n}$,
or equivalently $\tilde{n}=|\tilde{\Gamma}|/L^d$, we have
\begin{align}
&\ln\sum_{\{n_p\}}\theta\left(\sum_pH_{pp}^{(0)}+V^{(\delta,l)}_{\tilde{\gamma}}\leq|\tilde{\Gamma}|\varepsilon +\sum_{p<q}E_{pq}^{(0)}\right)
\nonumber \\
&\leq \delta^{-d}\left[ (n_{\rm sum}+\tilde{n}+\delta^d)\ln (n_{\rm sum}+\tilde{n}+\delta^d)
-n_{\rm sum}\ln n_{\rm sum} -\tilde{n}\ln\tilde{n} -\delta^d\ln\delta^d\right].
\end{align}
For sufficiently large $l$ and sufficiently small $\delta$, $|\tilde{\Gamma}|\approx L^d$ and therefore
\beq
n_{\rm sum}\approx \frac{1}{\delta\varepsilon}\left(\varepsilon-\varepsilon_g^{(0)}-\delta\varepsilon +\frac{\sigma_{\rm max}^2v}{2}\right)
\eeq
and $\tilde{n}\approx 1$.
Hence there is some positive constant $D=O(\ln(\delta\varepsilon)/\delta\varepsilon)$ independent of $\delta$ and $l$ with
\beq
\ln\sum_{\{n_p\}}\theta\left(\sum_pH_{pp}^{(0)}+V^{(\delta,l)}_{\tilde{\gamma}}\leq|\tilde{\Gamma}|\varepsilon +\sum_{p<q}E_{pq}^{(0)}\right)
\leq \delta^{-d}D
\label{eq:bound_nsum}
\eeq
for sufficiently large $l$ and sufficiently small $\delta$.
These restrictions of the summation over $k_p$ and $n_p$ are important for the evaluation of the upper bound.

\subsection*{Evaluation of a lower bound}
By using the left part of the inequality~(\ref{eq:ineq_H}), we have
\begin{align}
\theta\left(H^{(\delta,l)}_{\tilde{\gamma}}\leq|\tilde{\Gamma}|\varepsilon\right)
\geq& \sum_{\{n_p\}}\theta\left(\sum_pH_{pp}^{(0)}+V^{(\delta,l)}_{\tilde{\gamma}} \leq |\tilde{\Gamma}|\varepsilon - \sum_{p<q}E_{pq}^{(0)}\right)
\nonumber \\
&\times\prod_p\theta\left(\frac{H_{pp}^{(0)}}{l^d} - \varepsilon_g^{(0)} \in[l^dn_p\delta\varepsilon,l^d(n_p+1)\delta\varepsilon)\right)
\nonumber \\
\geq&\sum_{\{n_p\}}\theta\left( |\tilde{\Gamma}|\varepsilon_g^{(0)} + l^d\delta\varepsilon\sum_p(n_p+1) + V^{(\delta,l)}_{\tilde{\gamma}} \leq |\tilde{\Gamma}|\varepsilon - \sum_{p<q}E_{pq}^{(0)}\right)
\nonumber \\
&\times\prod_p\theta\left(\frac{H_{pp}^{(0)}}{l^d} - \varepsilon_g^{(0)} \in[l^dn_p\delta\varepsilon,l^d(n_p+1)\delta\varepsilon)\right)
\label{eq:lower_H}
\end{align}
We also have
\begin{align}
\theta\left(\frac{1}{|\tilde{\Gamma}|}\sum_{\bm{r}\in\hat{\tilde{\Gamma}}}\sigma(\bm{r}) \in [m,m+\delta m)\right)
=&\sum_{\{k_p\}}\theta\left(\frac{1}{|\tilde{\Gamma}|}\sum_pM_p\in [m,m+\delta m)\right)
\nonumber \\
&\times\prod_p\theta\left(\frac{M_p}{l^d}\in \left[k_p\frac{\delta m}{2},(k_p+1)\frac{\delta m}{2}\right)\right)
\nonumber \\
\geq&\sum_{\{k_p\}}\theta\left(|\tilde{\Gamma}|m \leq l^d\sum_pk_p\frac{\delta m}{2} < |\tilde{\Gamma}|(m+\delta m) - \frac{l^d}{2}\sum_p\delta m\right)
\nonumber \\
&\times\prod_p\theta\left(\frac{M_p}{l^d}\in \left[k_p\frac{\delta m}{2},(k_p+1)\frac{\delta m}{2}\right)\right).
\end{align}
By using $\sum_p1=|\tilde{\Gamma}|/l^d$, we obtain
\begin{align}
\theta\left(\frac{1}{|\tilde{\Gamma}|}\sum_{\bm{r}\in\hat{\tilde{\Gamma}}}\sigma(\bm{r}) \in [m,m+\delta m)\right)
\geq\sum_{\{k_p\}}\theta\left(|\tilde{\Gamma}|m \leq l^d\sum_pk_p\frac{\delta m}{2} < |\tilde{\Gamma}|\left(m+\frac{\delta m}{2}\right)\right)
\nonumber \\
\times\prod_p\theta\left(\frac{M_p}{l^d}\in \left[k_p\frac{\delta m}{2},(k_p+1)\frac{\delta m}{2}\right)\right).
\label{eq:lower_M}
\end{align}

From Eqs.~(\ref{eq:lower_H}) and (\ref{eq:lower_M}), the lower bound of the coarse-grained entropy given in Eq.~(\ref{eq:cg_s2}) is obtained:
\begin{align}
&s^{(\delta,l)}(\varepsilon,m,\delta m;\tilde{\Gamma})
\geq 
\nonumber \\
&\frac{1}{|\tilde{\Gamma}|} \ln \sum_{\{n_p\}}\sum_{\{k_p\}}
\theta\left( |\tilde{\Gamma}|\varepsilon_g^{(0)} + l^d\delta\varepsilon\sum_p(n_p+1) + V^{(\delta,l)}_{\tilde{\gamma}} \leq |\tilde{\Gamma}|\varepsilon - \sum_{p<q}E_{pq}^{(0)}\right)\
\nonumber \\
&\times\theta\left(|\tilde{\Gamma}|m\leq l^d\sum_pk_p\frac{\delta m}{2}<|\tilde{\Gamma}|\left(m+\frac{\delta m}{2}\right)\right)
\nonumber \\
&\times \prod_p\left[\sum_{\bm{\sigma}_{\Lambda_l^{(p)}}}
\theta\left(\frac{H_{pp}^{(0)}}{l^d}-\varepsilon_g^{(0)}\in[n_p\delta\varepsilon,(n_p+1)\delta\varepsilon)\right)
\theta\left(\frac{M_p}{l^d}\in\left[k_p\frac{\delta m}{2},(k_p+1)\frac{\delta m}{2}\right)\right)\right].
\end{align}
In terms of the entropy density of the reference system, 
\beq
s^{(0)}(\varepsilon,m,\delta m;\Lambda_l^{(p)}) = \frac{1}{l^d}\ln \sum_{\bm{\sigma}_{\Lambda_l^{(p)}}}\theta(H_{pp}^{(0)}\leq l^d\varepsilon) \theta\left(\frac{M_p}{l^d}\in[m,m+\delta m)\right),
\eeq
we have
\begin{align}
s^{(\delta,l)}(\varepsilon,m,\delta m;\tilde{\Gamma})
\geq&\frac{1}{|\tilde{\Gamma}|} \ln \sum_{\{n_p\}}\sum_{\{k_p\}}
\theta\left( |\tilde{\Gamma}|\varepsilon_g^{(0)} + l^d\delta\varepsilon\sum_p(n_p+1) + V^{(\delta,l)}_{\tilde{\gamma}} \leq |\tilde{\Gamma}|\varepsilon - \sum_{p<q}E_{pq}^{(0)}\right)\
\nonumber \\
&\times\theta\left(|\tilde{\Gamma}|m\leq l^d\sum_pk_p\frac{\delta m}{2}<|\tilde{\Gamma}|\left(m+\frac{\delta m}{2}\right)\right)
\nonumber \\
&\times \prod_p\left\{
\exp\left[l^ds^{(0)}\left(\varepsilon_g^{(0)}+(n_p+1)\delta\varepsilon,k_p\frac{\delta m}{2},\frac{\delta m}{2};\Lambda_l^{(p)}\right)\right]\right.
\nonumber \\
&\qquad \left. -\exp\left[l^ds^{(0)}\left(\varepsilon_g^{(0)}+n_p\delta\varepsilon,k_p\frac{\delta m}{2},\frac{\delta m}{2};\Lambda_l^{(p)}\right)\right]\right\}
\end{align}
By keeping only the term with $k_p$ and $n_p$ which maximize the quantity inside logarithm,
we obtain
\begin{align}
s^{(\delta,l)}(\varepsilon,m,\delta m;\tilde{\Gamma})
\geq
\max_{\{n_p\},\{k_p\}}&\left[\frac{l^d}{|\tilde{\Gamma}|} \sum_p s^{(0)}\left(\varepsilon_g^{(0)}+(n_p+1)\delta\varepsilon,k_p\frac{\delta m}{2},\frac{\delta m}{2};\Lambda_l^{(p)}\right)\right.
\nonumber \\
&\qquad -\frac{1}{|\tilde{\Gamma}|}\sum_p\left(1-e^{-l^d\delta s}\right)
\nonumber \\
&: |\tilde{\Gamma}|\varepsilon_g^{(0)}+l^d\delta\varepsilon\sum_p(n_p+1)+V^{(\delta,l)}_{\tilde{\gamma}}\leq |\tilde{\Gamma}|\varepsilon-\sum_{p<q}E_{pq}^{(0)},
\nonumber \\
&\left. \quad |\tilde{\Gamma}|m\leq l^d\frac{\delta m}{2}\sum_pk_p < |\tilde{\Gamma}|\left(m+\frac{\delta m}{2}\right)\right],
\end{align}
where 
$$\delta s=s^{(0)}\left(\varepsilon_g^{(0)}+(n_p+1)\delta\varepsilon,k_p\frac{\delta m}{2},\frac{\delta m}{2};\Lambda_l^{(p)}\right)-s^{(0)}\left(\varepsilon_g^{(0)}+n_p\delta\varepsilon,k_p\frac{\delta m}{2},\frac{\delta m}{2};\Lambda_l^{(p)}\right).$$
In the limit of $l\rightarrow\infty$ and then $\delta m\rightarrow 0$ and $\delta\varepsilon\rightarrow 0$, by the assumption that the entropy density of the reference system has a thermodynamic limit, Eq.~(\ref{eq:limit_reference_s}), we obtain
\begin{align}
&\liminf_{\delta m\rightarrow 0 \text{ after }l\rightarrow\infty}s^{(\delta,l)}(\varepsilon,m,\delta m;\tilde{\Gamma})
\nonumber \\
&\geq \max_{\{\varepsilon_p,m_p\}}\left[\frac{\delta^d}{|\tilde{\gamma}|}\sum_ps^{(0)}(\varepsilon_p,m_p):\delta^d\sum_p\varepsilon_p+v_{\tilde{\gamma}}^{(\delta)}\leq|\tilde{\gamma}|\varepsilon,
\delta^d\sum_pm_p=|\tilde{\gamma}|m\right],
\end{align}
where $v^{(\delta)}_{\tilde{\gamma}}=\lim_{l\rightarrow\infty}V^{(\delta,l)}_{\tilde{\gamma}}/L^d$.

Furthermore, we shall take the limit of $\delta\rightarrow 0$.
Now we define the domain $\lambda_{\delta}^{(p)}=\Lambda_l^{(p)}/L=\delta(\Lambda_l^{(p)}/l)$.
The domain $\lambda_{\delta}^{(p)}$ is independent of $l$ and $|\lambda_{\delta}^{(p)}|=\delta^d$.
The central position of $\lambda_{\delta}^{(p)}$ is denoted by $\bm{x}_p$.
By introducing the functions $\varepsilon'(\bm{x})$ and $m'(\bm{x})$ so that $\varepsilon'(\bm{x})=\varepsilon_p$ and $m'(\bm{x})=m_p$ for any $\bm{x}\in\lambda_{\delta}^{(p)}$, we can write
\beq
\begin{aligned}
&\delta^d\sum_ps^{(0)}(\varepsilon_p,m_p)=\int_{\tilde{\gamma}}d^d\bm{x}s^{(0)}(\varepsilon'(\bm{x}),m'(\bm{x})),
\\
&\delta^d\sum_p\varepsilon_p=\int_{\tilde{\gamma}}d^d\bm{x}\varepsilon'(\bm{x}),
\\
&\delta^d\sum_pm_p=\int_{\tilde{\gamma}}d^d\bm{x}m'(\bm{x}),
\\
&\tilde{v}_{\tilde{\gamma}}^{(\delta)}=-\frac{1}{2}\int_{\tilde{\gamma}}d^d\bm{x} \int_{\tilde{\gamma}}d^d\bm{y} \phi(\bm{x},\bm{y})m'(\bm{x})m'(\bm{y}).
\label{eq:step}
\end{aligned}
\eeq
In the limit of $\delta\rightarrow 0$, $\tilde{\gamma}\rightarrow\gamma$ and any Riemann integrable functions $\varepsilon(\bm{x})$ and $m(\bm{x})$ can be approximated by step functions $\varepsilon'(\bm{x})$ and $m'(\bm{x})$ if $\{\varepsilon_p\}$ and $\{m_p\}$ are suitably chosen.
We therefore obtain
\begin{align}
&\liminf_{\delta\rightarrow 0}\liminf_{\delta m\rightarrow 0 \text{ after } l\rightarrow\infty}s^{(\delta,l)}(\varepsilon,m,\delta m;\tilde{\Gamma})
\nonumber \\
&\geq \sup_{\varepsilon(\cdot),m(\cdot)\in\mathcal{R}_{\gamma}}\left[\int_{\gamma}d^d\bm{x}s^{(0)}(\varepsilon(\bm{x}),m(\bm{x}))
: \int_{\gamma}d^d\bm{x}m(\bm{x})=m,
\right. \nonumber \\ 
&\left.
\qquad\int_{\gamma}d^d\bm{x}\varepsilon(\bm{x})-\frac{1}{2}\int_{\gamma}d^d\bm{x}\int_{\gamma}d^d\bm{y}\phi(\bm{x},\bm{y})m(\bm{x})m(\bm{y})\leq\varepsilon \right]
\end{align}

\subsection*{Evaluation of an upper bound}

By using the right part of Eq.~(\ref{eq:ineq_H}), we have
\begin{align}
\theta\left(H^{(\delta,l)}_{\tilde{\gamma}}\leq|\tilde{\Gamma}|\varepsilon\right)
\leq \sum_{\{n_p\}}\theta\left(|\tilde{\Gamma}|\varepsilon_g^{(0)} + l^d\delta\varepsilon\sum_pn_p + V^{(\delta,l)}_{\tilde{\gamma}} \leq |\tilde{\Gamma}|\varepsilon + \sum_{p<q}E_{pq}^{(0)} \right)
\nonumber \\
\times\prod_p \theta\left( \frac{H_{pp}^{(0)}}{l^d}-\varepsilon_g^{(0)} \in [n_p\delta\varepsilon,(n_p+1)\delta\varepsilon)\right)
\label{eq:upper_H}
\end{align}
and
\begin{align}
\theta\left(\frac{1}{|\tilde{\Gamma}|}\sum_{\bm{r}\in\hat{\tilde{\Gamma}}}\sigma(\bm{r}) \in [m,m+\delta m) \right) \leq \sum_{\{k_p\}}\theta\left( |\tilde{\Gamma}|\left(m-\frac{\delta m}{2}\right) \leq l^d\sum_pk_p\frac{\delta m}{2} < |\tilde{\Gamma}|(m+\delta m) \right)
\nonumber \\
\times\prod_p \theta\left( \frac{M_p}{l^d}\in\left[k_p\frac{\delta m}{2},(k_p+1)\frac{\delta m}{2}\right) \right).
\label{eq:upper_M}
\end{align}
Substituting Eqs.~(\ref{eq:upper_H}) and (\ref{eq:upper_M}) into Eq.~(\ref{eq:cg_s2}), we obtain
\begin{align}
&s^{(\delta,l)}(\varepsilon,m,\delta m;\tilde{\Gamma})
\nonumber \\
&\leq \frac{1}{|\tilde{\Gamma}|} \ln \sum_{\{n_p\}} \sum_{\{k_p\}}
\theta\left(|\tilde{\Gamma}|\varepsilon_g^{(0)} + l^d\delta\varepsilon\sum_pn_p + V^{(\delta,l)}_{\tilde{\gamma}} \leq |\tilde{\Gamma}|\varepsilon + \sum_{p<q}E_{pq}^{(0)} \right)
\nonumber \\
&\times\theta\left( |\tilde{\Gamma}|\left(m-\frac{\delta m}{2}\right) \leq l^d\sum_pk_p\frac{\delta m}{2} < |\tilde{\Gamma}|(m+\delta m) \right)
\nonumber \\
&\times\prod_p \left[ \sum_{\bm{\sigma}_{\Lambda_l^{(p)}}}
\theta\left( \frac{H_{pp}^{(0)}}{l^d}-\varepsilon_g^{(0)} \in [n_p\delta\varepsilon,(n_p+1)\delta\varepsilon)\right)
\theta\left( \frac{M_p}{l^d}\in\left[k_p\frac{\delta m}{2},(k_p+1)\frac{\delta m}{2}\right) \right) \right].
\end{align}
In terms of the entropy density of the reference system, the above inequality is written as
\begin{align}
s^{(\delta,l)}(\varepsilon,m,\delta m;\tilde{\Gamma})
\leq
\frac{1}{|\tilde{\Gamma}|} \ln \sum_{\{n_p\}} \sum_{\{k_p\}}
\theta\left(|\tilde{\Gamma}|\varepsilon_g^{(0)} + l^d\delta\varepsilon\sum_pn_p + V^{(\delta,l)}_{\tilde{\gamma}} \leq |\tilde{\Gamma}|\varepsilon + \sum_{p<q}E_{pq}^{(0)} \right)
\nonumber \\
\times\theta\left( |\tilde{\Gamma}|\left(m-\frac{\delta m}{2}\right) \leq l^d\sum_pk_p\frac{\delta m}{2} < |\tilde{\Gamma}|(m+\delta m) \right)
\nonumber \\
\times\prod_p \left\{ \exp\left[l^ds^{(0)}\left(\varepsilon_g^{(0)}+(n_p+1)\delta\varepsilon,k_p\frac{\delta m}{2},\frac{\delta m}{2};\Lambda_l^{(p)}\right)\right] \right.
\nonumber \\
\left.- \exp\left[l^ds^{(0)}\left(\varepsilon_g^{(0)}+n_p\delta\varepsilon,k_p\frac{\delta m}{2},\frac{\delta m}{2};\Lambda_l^{(p)}\right)\right]\right\}.
\end{align}
Because $k_{\rm min}\leq k_p\leq k_{\rm max}$,
\beq
\sum_{\{k_p\}}f(\{k_p\})\leq \left(\sum_{\{k_p\}}1\right)\max_{\{k_p\}}f(\{k_p\})\leq\left(\frac{4\sigma_{\max}}{\delta m}+2\right)\max_{\{k_p\}}f(\{k_p\}).
\eeq
Moreover, because of Eq.~(\ref{eq:bound_nsum}),
\beq
\sum_{\{n_p\}}f(\{n_p\})\leq e^{\delta^{-d}D}\max_{\{n_p\}}f(\{n_p\}),
\eeq
where the summation over $\{n_p\}$ is restricted so that $\sum_pH_{pp}^{(0)}+V^{(\delta,l)}_{\tilde{\gamma}}\leq|\tilde{\Gamma}|\varepsilon +\sum_{p<q}E_{pq}^{(0)}$.
An upper bound of the entropy is thus given by
\begin{align}
&s^{(\delta,l)}(\varepsilon,m,\delta m;\tilde{\Gamma})
\nonumber \\
&\leq \max_{\{n_p\},\{k_p\}}\left[ \frac{l^d}{|\tilde{\Gamma}|} \sum_p s^{(0)}\left(\varepsilon_g^{(0)}+(n_p+1)\delta\varepsilon,k_p\frac{\delta m}{2},\frac{\delta m}{2};\Lambda_l^{(p)}\right) -\frac{1}{|\tilde{\Gamma}|}\left(1-e^{-l^d\delta s}\right) \right.
\nonumber \\
&\qquad\qquad\qquad\qquad: |\tilde{\Gamma}|\varepsilon_g^{(0)}+l^d\delta\varepsilon\sum_pn_p + V^{(\delta,l)}_{\tilde{\gamma}} \leq |\tilde{\Gamma}|\varepsilon + \sum_{p<q}E_{pq}^{(0)},
\nonumber \\
&\qquad\qquad\qquad\qquad\left.|\tilde{\Gamma}|\left(m-\frac{\delta m}{2}\right)\leq l^d\sum_pk_p\frac{\delta m}{2} < |\tilde{\Gamma}|(m+\delta m)\right]
\nonumber \\
&+\frac{1}{|\tilde{\Gamma}|}\ln\left(\frac{4\sigma_{\rm max}}{\delta m}+2\right) +\frac{1}{|\tilde{\Gamma}|}\delta^{-d}D.
\end{align}
In the limit of $l\rightarrow\infty$ and then $\delta m\rightarrow 0$ and $\delta\varepsilon\rightarrow 0$, we obtain
\begin{align}
\limsup_{\delta m\rightarrow 0 \text{ after }l\rightarrow\infty}s^{(\delta,l)}(\varepsilon,m,\delta m;\tilde{\Gamma})
\leq \max_{\{\varepsilon_p\},\{m_p\}}\left[\frac{\delta^d}{|\tilde{\gamma}|}\sum_ps^{(0)}(\varepsilon_p,m_p)
: \delta^d\sum_pm_p=|\tilde{\gamma}|m,
\right. \nonumber \\
\left. 
\delta^d\sum_p\varepsilon_p +v^{\delta}_{\tilde{\gamma}} \leq |\tilde{\gamma}|\varepsilon\right].
\end{align}
Here we again introduce the step functions $\varepsilon'(\bm{x})$ and $m'(\bm{x})$.
Since $\varepsilon'(\bm{x})$ and $m'(\bm{x})$ are Riemann integrable, we have, by using Eq.~(\ref{eq:step}),
\begin{align}
&\limsup_{\delta m\rightarrow 0 \text{ after }l\rightarrow\infty}s^{(\delta,l)}(\varepsilon,m,\delta m;\tilde{\Gamma})
\nonumber \\
&\leq\max_{\{\varepsilon_p\},\{m_p\}}\left[ \frac{1}{|\tilde{\gamma}|}\int_{\tilde{\gamma}}d^d\bm{x}s^{(0)}(\varepsilon'(\bm{x}),m'(\bm{x}))
: \int_{\tilde{\gamma}}d^d\bm{x}m'(\bm{x})=|\tilde{\gamma}|m,\right.
\nonumber \\
&\qquad\qquad\qquad\left. \int_{\tilde{\gamma}}d^d\bm{x}\varepsilon'(\bm{x})-\frac{1}{2}\int_{\tilde{\gamma}}\int_{\tilde{\gamma}} \phi(\bm{x},\bm{y}) m'(\bm{x})m'(\bm{y}) \leq |\tilde{\gamma}\varepsilon \right]
\nonumber \\
&\leq \sup_{\varepsilon(\cdot),m(\cdot)\in\mathcal{R}_{\gamma}}\left[ \frac{1}{|\tilde{\gamma}|}\int_{\tilde{\gamma}}d^d\bm{x}s^{(0)}(\varepsilon(\bm{x}),m(\bm{x}))
: \int_{\tilde{\gamma}}d^d\bm{x}m(\bm{x})=|\tilde{\gamma}|m,\right.
\nonumber \\
&\qquad\qquad\qquad\left. \int_{\tilde{\gamma}}d^d\bm{x}\varepsilon(\bm{x})-\frac{1}{2}\int_{\tilde{\gamma}}\int_{\tilde{\gamma}} \phi(\bm{x},\bm{y}) m(\bm{x})m(\bm{y}) \leq |\tilde{\gamma}|\varepsilon \right].
\end{align}
In the limit of $\delta\rightarrow 0$, $\tilde{\gamma}\rightarrow\gamma$ and thus
\begin{align}
&\limsup_{\delta\rightarrow 0}\limsup_{\delta m\rightarrow 0 \text{ after } l\rightarrow\infty} s^{(\delta,l)}(\varepsilon,m,\delta m;\tilde{\Gamma})
\nonumber \\
&\leq
\sup_{\varepsilon(\cdot),m(\cdot)\in\mathcal{R}_{\gamma}}\left[ \int_{\gamma}d^d\bm{x}s^{(0)}(\varepsilon(\bm{x}),m(\bm{x}))
: \int_{\gamma}d^d\bm{x}m(\bm{x})=m, \right.
\nonumber \\
&\left.\qquad\qquad\qquad\int_{\gamma}d^d\bm{x}\varepsilon(\bm{x})-\frac{1}{2}\int_{\gamma}\int_{\gamma} \phi(\bm{x},\bm{y}) m(\bm{x})m(\bm{y}) \leq \varepsilon \right].
\end{align}
This is identical to the derived lower bound.
We therefore have finished to prove Lemma~\ref{lemma2}.

\section{Application of the variational formula to the case of periodic boundary conditions}
\label{sec:periodic}

We have considered spin systems with free boundary conditions, but Theorem~\ref{theorem} also holds for periodic boundary conditions as long as $|\bm{x}-\bm{y}|$ in the conditions~(\ref{eq:condition1}) and (\ref{eq:condition2}) is interpreted by the minimum image convention.
In this section, we briefly mention some consequences from the variatonal expression of the entropy density.

\subsection{Exactness of the mean-field theory and its violation}

In this section, we assume periodic boundary conditions and fully ferromagnetic and translational invariant couplings $\phi(\bm{x},\bm{y})=\phi(\bm{x}-\bm{y})\geq 0$ for all $\bm{x}$ and $\bm{y}$, we can obtain some results mentioned below.

In periodic boundary conditions, we set $\gamma$ to be the $d$-dimensional unit cube, $\gamma=[0,1)^d\equiv\Lambda_1$.
In general, we define $\Lambda_l^d\equiv [0,l)^d$, that is, $\Lambda_l^d$ is the $d$-dimensional cube of side $l$.
The point $\bm{x}\pm\bm{e}_k$ is identified with $\bm{x}$, where $\bm{e}_k$ is the unit vector along $k$-direction ($k=1,2,\dots, d$).
Of course, the translational invariant potential satisfies $\phi(\bm{x})=\phi(\bm{x}\pm\bm{e}_k)$ for $k=1,2,\dots, d$.

In the above setting, it is shown that the interaction potential $\phi(\bm{x}-\bm{y})$ can be replaced by the mean-field (MF) coupling, $\phi(\bm{x}-\bm{y})\rightarrow 1$ for all $\bm{x},\bm{y} \in\Lambda_1$ in a wide region of the parameter space, $(\varepsilon,m)$ or $(\beta, m)$ depending on the ensemble, called the ``MF region'' without changing the value of the entropy density~\cite{Mori2012_microcanonical} or the free energy density~\cite{Mori2010_analysis,Mori2011_instability}.
On the other hand, if the density of an extensive quantity such as $\varepsilon$ or $m$ is held fixed, it is also shown that there is a parameter region called the non-MF region, in which the value of the entropy density or the free energy density crucially depends on the details of $\phi(\bm{x}-\bm{y})$, and hence replacing $\phi$ by 1 is not allowed.
The fact that replacing $\phi$ by 1 is allowed is called the ``exactness of the MF theory''~\cite{Cannas-Tamarit1996,Cannas2000,Campa2000,Campa2003}, because it is well known that the spin model with the all-to-all couplings are thermodynamically equivalent to the spin model with the MF approximation~\cite{Nishimori_text}.

In earlier works~\footnote
{Exactness of the MF theory and its violation has been also discussed for quantum spin systems~\cite{Mori2012_equilibrium}, but the derivation of the {\it microcanonical} entropy in quantum systems has not been fully rigorous as pointed out by Olivier and Kastner~\cite{Olivier-Kastner2014}.
However, the results discussed in classical spins are also true in quantum spin systems at least for the {\it canonical} ensemble.},
the exactness of the MF theory and its violation has been investigated for the case with the homogeneous magnetic field but without any short-range interactions, $H_{\Gamma}^{(0)}=-h\sum_{\bm{r}\in\hat{\Gamma}}\sigma(\bm{r})$.
As we will show below, the results of the earlier works are straightforwardly extended to the case with short-range interactions by using the variatonal expression of the entropy density~(\ref{eq:theorem}).

We shall derive the exactness of the MF theory and its violation for the canonical ensemble.
In periodic boundary conditions, the translationally symmetric potential energy can be diagonalized by the Fourier expansion,
\beq
\int_{\Lambda_1}d^d\bm{x}\int_{\Lambda_1}d^d\bm{y}\phi(\bm{x}-\bm{y})m(\bm{x})m(\bm{y})
=\sum_{\bm{n}\in\mathbb{Z}^d}\phi_{\bm{n}}|m_{\bm{n}}|^2,
\eeq
where
\beq
\phi_{\bm{n}}=\int_{\Lambda_1}d^d\bm{x}\phi(\bm{x})e^{-2\pi i\bm{n}\cdot\bm{x}}
=\int_{\Lambda_1}d^d\bm{x}\phi(\bm{x})\cos(2\pi\bm{n}\cdot\bm{x})
\label{eq:int_Fourier}
\eeq
and
\beq
m_{\bm{n}}=\int_{\Lambda_1}d^d\bm{x}m(\bm{x})e^{2\pi i\bm{n}\cdot\bm{x}}.
\eeq
From Eq.~(\ref{eq:int_Fourier}), as long as $\phi(\bm{x})\geq 0$ for any $\bm{x}\in\Lambda_1$,
\beq
\phi_{\bm{n}}\leq\int_{\Lambda_1}d^d\bm{x}\phi(\bm{x})=\phi_0=1.
\label{eq:int_bound}
\eeq
Remember the normalization of $\mathcal{N}_{\phi,\gamma}=1$ in Eq.~(\ref{eq:integrable}).

We define the second largest Fourier component of $\phi(\bm{x})$ as $\phi_{\rm max}$,
\beq
\phi_{\rm max}=\max_{\bm{n}\in\mathbb{Z}^d\backslash 0}\phi_{\bm{n}}.
\label{eq:phi_max}
\eeq
The interaction term is bounded as~\cite{Mori2013_statphys}
\beq
\sum_{\bm{n}\in\mathbb{Z}^d}\phi_{\bm{n}}|m_{\bm{n}}|^2
\leq  m^2 + \phi_{\rm max} \int_{\Lambda_1}d^d\bm{x} m(\bm{x})^2 - \phi_{\rm max}m^2.
\eeq
By using this inequality, we find that the free energy density satisfies
\begin{align}
f(\beta,m) \geq&\inf_{m(\cdot)\in\mathcal{R}_{\Lambda_1}}
\left\{ -\frac{1}{2}m^2+f^{(0)}(\beta,m) + \left[ \int_{\Lambda_1}d^d\bm{x}\left( -\frac{\phi_{\rm max}}{2}m(\bm{x})^2+f^{(0)}(\beta,m(\bm{x}))\right)
\right.\right. \nonumber \\ 
&\left.\left. \qquad\qquad
-\left(-\frac{\phi_{\rm max}}{2}m^2+f^{(0)}(\beta,m)\right) \right] \right\}.
\end{align}
We define the free energy of the reference system with the MF couplings as
\beq
f_{\rm MF}(\beta,m;J)=-\frac{J}{2}m^2+f^{(0)}(\beta,m).
\eeq
Then the lower bound of the free energy is written as
\beq
f(\beta,m) \geq f_{\rm MF}(\beta,m;1) - \left[ f_{\rm MF}(\beta,m;\phi_{\rm max}) - f_{\rm MF}^{**}(\beta,m;\phi_{\rm max})\right],
\label{eq:free_lower}
\eeq
where $f_{\rm MF}^{**}(\beta,m;\phi_{\rm max})$ is the convex envelope of $f_{\rm MF}(\beta,m;\phi_{\rm max})$ with respect to $m$.
In other words, $f_{\rm MF}^{**}(\beta,m;\phi_{\rm max})$ is the maximum convex function of $m$ satisfying $f_{\rm MF}^{**}(\beta,m)\leq f_{\rm MF}(\beta,m;\phi_{\rm max})$.
We have used the relation
\beq
f_{\rm MF}^{**}(\beta,m;\phi_{\rm max})=
\inf_{m(\cdot)\in\mathcal{R}_{\Lambda_1}}\left[ \int_{\Lambda_1}d^d\bm{x} f_{\rm MF}(\beta,m(\bm{x});\phi_{\rm max}) : \int_{\Lambda_1}d^d\bm{x}m(\bm{x})=m\right].
\eeq

The upper bound of the free energy density is easily obtained by putting $m(\bm{x})=m$ in Eq.~(\ref{eq:theorem_free}),
\beq
f(\beta,m)\leq -\frac{1}{2}m^2 + f^{(0)}(\beta,m) = f_{\rm MF}(\beta,m;1).
\eeq
Thus we have obtained the following inequality:
\beq
f_{\rm MF}(\beta,m;1)-\left[ f_{\rm MF}(\beta,m;\phi_{\rm max})-f_{\rm MF}^{**}(\beta,m;\phi_{\rm max}) \right] \leq f(\beta,m) \leq f_{\rm MF}(\beta,m;1).
\label{eq:free_inequality}
\eeq
This inequality is an extension of the inequality derived in the previous work~\cite{Mori2010_analysis,Mori2011_instability}, in which only the case of $H^{(0)}_{\Gamma}=-h\sum_{\bm{r}\in\hat{\Gamma}}\sigma(\bm{r})$, i.e., without short-range interactions, was considered\footnote
{For a reference Hamiltonian of the form $H^{(0)}_{\Gamma}=-h\sum_{\bm{r}\in\hat{\Gamma}}\sigma(\bm{r})$, $f_{\rm MF}(\beta,m;\phi_{\rm max})=f_{\rm MF}(\beta\phi_{\rm max},m;1)$.
If we write $f_{\rm MF}(\beta,m)\equiv f_{\rm MF}(\beta,m;1)$, the inequality~(\ref{eq:free_inequality}) is reduced to the inequality obtained in Ref.~\cite{Mori2010_analysis,Mori2011_instability}.}.
From the inequality~(\ref{eq:free_inequality}), if $f_{\rm MF}(\beta,m;\phi_{\rm max})$ is convex with respect to $m$, the lower bound coincides with the upper bound, and thus $f(\beta,m)=f_{\rm MF}(\beta,m;1)$.
In particular, at the minimum point of $f_{\rm MF}(\beta,m;1)$ with respect to $m$, which corresponds to an equilibrium state when the value of the magnetization is not fixed, the convexity of $f_{\rm MF}(\beta,m;\phi_{\rm max})$ is always satisfied and thus $\min_mf(\beta,m)=\min_mf_{\rm MF}(\beta,m;1)$.
This is nothing but the statement of the exactness of the MF theory.

We shall consider the case where $m$ is held fixed at some value, not necessarily the minimum of $f(\beta,m)$.
From the stability analysis around the uniform solution $m(\bm{x})=m$, it is found that in the region of the parameter space $(\beta,m)$ with $\d^2f_{\rm MF}(\beta,m;\phi_{\rm max})/\d m^2<0$, the uniform solution corresponds to a local maximum point of the free energy functional
\beq
\mathcal{F}(\beta,\{ m(\bm{x})\}) = -\frac{1}{2}\int_{\Lambda_1}d^d\bm{x} \int_{\Lambda_1}d^d\bm{y}\phi(\bm{x}-\bm{y})m(\bm{x})m(\bm{y}) + \int_{\Lambda_1}d^d\bm{x} f^{(0)}(\beta,m(\bm{x})),
\eeq
which means that the uniform solution is unstable.
We therefore have
\beq
f(\beta,m)<f_{\rm MF}(\beta,m;1)
\eeq
for $(\beta,m)$ satisfying $\d^2f_{\rm MF}(\beta,m;\phi_{\rm max})/\d m^2<0$.
This is a part of the non-MF region.
In the non-MF region, macroscopic heterogeneity emerges, see Ref.~\cite{Mori2011_instability,Mori2013_statphys} in more detail.

\subsection{Mean-field universality for critical phenomena}

In short-range interacting systems, large clusters with the same spin state appear near the critical point, which implies the divergence of the correlation length~\cite{Nishimori_text}.
The universality class depends on the type of symmetry breaking, spacial dimension, and so on.
On the other hand, in long-range interacting systems, the system tends to be homogeneous even at the critical point because all the spins interact with each other and spacial geometry becomes less important.
Indeed, in the model only with the all-to-all interactions (the MF model), the spin configuration is always uniform and the universality class of the critical phenomena belong to the MF universality class independently of the spacial dimension.

When the system possesses both the short- and long-range interactions, it has been argued that critical phenomena always belong to the MF universality class even if the strength of long-range couplings is infinitesimal~\cite{Nakada2011}.
We can see it by using the exactness of the MF theory.
When the temperature is above the critical temperature, the free energy is convex with respect to $m$ and $f(\beta,m)=f_{\rm MF}(\beta,m;1)=-(1/2)m^2+f^{(0)}(\beta,m)$.
We assume that $m=0$ is the minimum point of $f(\beta,m)$.
If there were no long-range interaction, the macroscopic ordering due to short-range interactions would occur at $\beta_c^{(0)}$ with $\d^2f^{(0)}(\beta_c^{(0)},m)/\d m^2|_{m=0}=0$.
With the presence of long-range interactions, at the critical inverse temperature $\beta_c$, $\d^2f(\beta_c,m)/\d m^2|_{m=0}=-1+\d^2f^{(0)}(\beta_c,m)/\d m^2|_{m=0}=0$, which implies that $\d^2f^{(0)}(\beta_c,m)/\d m^2|_{m=0}=1>0$.
From this observation, we can say $\beta_c<\beta_c^{(0)}$ and phase transitions in a system with short- and long-range interactions are always driven by long-range interactions before growing large clusters due to short-range interactions.
As a result, critical phenomena are governed by long-range interactions and the critical phenomena belong to the MF universality class.
The crossover between short-range Ising model and the long-range Ising model is investigated in Ref.~\cite{Nakada2011}.

\section{Discussion}
\label{sec:discussion}

We have proven the existence of the thermodynamic limit in spin systems where both short-range interactions and long-range ones are present.
We have obtained the variational expression of the entropy density explicitly depending on the shape of the system $\gamma$.
This implies nonadditivity, which is one of the important characteristics of long-range interacting systems~\cite{Campa_review2009}.

The variational expression leads us to some results such as the exactness of the MF theory and the MF universality class of critical phenomena in systems where short-range interactions compete with long-range interactions.

In the proof of Lemma~\ref{lemma1}, we have evaluated the difference between the original Hamiltonian and the coarse-grained one.
We divide it into $\Delta_1$ and $\Delta_2$, the former of which stands for the contribution from the short-distance fluctuations and the latter of which represents the contribution from the long-distance fluctuations.
By comparing Eq.~(\ref{eq:Delta1}) with Eqs.~(\ref{eq:Delta2a}), (\ref{eq:Delta2b}), and (\ref{eq:Delta2c}), it is obvious that $\Delta_1\ll\Delta_2$ for $\alpha\leq d-1$ and $\Delta_1\approx\Delta_2$ for $d-1<\alpha<d$ when $\delta$ is very small.
This implies that the short-distance fluctuations around the coarse-grained magnetization $\{ m(\bm{x})\}$ become important only for $d-1<\alpha<d$.
Although the equilibrium state itself is qualitatively irrespective of whether $\alpha\leq d-1$ or $d-1<\alpha<d$, the nonequilibrium dynamics would be qualitatively different in the two regimes.
Indeed, the dynamical classification of long-range interactions has been suggested, where it was shown that $\alpha\leq d-1$ and $\alpha>d-1$ belong to the different classes~\cite{Gabrielli_prl2010,Gabrielli2010}.

In long-range interacting systems, not only equilibrium properties but also dynamical properties are peculiar, e.g., ergodicity breaking~\cite{Mukamel2005}, existence of quasi-stationary states~\cite{Antoni-Ruffo1995}, dynamical criticality near the spinodal point~\cite{Mori2010_asymptotic}, and so on, see Ref.~\cite{Levin_review2014} for a review.
Therefore, it is interesting to explore those dynamical properties for general cases, e.g., for arbitrary pair interactions under suitable conditions, for an arbitrary shape of the system, and for the case where both short-range interactions and long-range ones are present.
The coarse graining will be also applicable for some dynamical problems and a thermodynamic function will play the role of a dynamical potential~\cite{Mori2010_asymptotic}.
We therefore hope that this work will also serve as a guide to explore dynamical issues in long-range interacting systems.

\end{document}